\documentstyle[prb,aps,eqsecnum,multicol,epsf]{revtex}
\newcommand{\bleq}{\ifpreprintsty
    \else
   \end{multicols}\vspace*{-3.5ex}{\tiny
   \noindent\begin{tabular}[t]{c|}
   \parbox{0.493\hsize}{~} \\ \hline \end{tabular}}
    \fi}
\newcommand{\eleq}{\ifpreprintsty
    \else
   {\tiny\hspace*{\fill}\begin{tabular}[t]{|c}\hline
   \parbox{0.49\hsize}{~} \\
   \end{tabular}}\vspace*{-2.5ex}\begin{multicols}{2}
   \noindent
    \fi}
\newcommand{\bcols}{\ifpreprintsty\else\begin{multicols}{2}\fi}
\newcommand{\ecols}{\ifpreprintsty\else\end{multicols}\fi}
\begin{document}
\draft


\title{ 
 Effect of nearest- and next-nearest neighbor interactions 
 on the spin-wave velocity \\
 of one-dimensional quarter-filled spin-density-wave conductors 
  }

\author{
Y. Tomio$^{1,}$%
\footnote{E-mail: tomio@edu2.phys.nagoya-u.ac.jp} 
,  
N. Dupuis$^{2,}$%
\footnote{E-mail: dupuis@lps.u-psud.fr}
and 
Y. Suzumura$^{1,}$%
\footnote{E-mail: e43428a@nucc.cc.nagoya-u.ac.jp}
}

\address{
$^{1}$Department of Physics, Nagoya University, Nagoya 464-8602, Japan \\ 
$^{2}$Laboratoire de Physique des Solides, Associ\'e au CNRS, 
 Universit\'{e} Paris-Sud, 91405 Orsay, France \\
}

\date{\hspace*{20mm}}
\maketitle

\begin{abstract}
 We study spin fluctuations in quarter-filled one-dimensional
spin-density-wave systems in presence of short-range Coulomb
interactions. By applying a  path integral method, 
 the spin-wave velocity is calculated  as a function of on-site
($U$), nearest ($V$) and next-nearest ($V_2$) neighbor-site interactions.   
With increasing $V$ or $V_2$, the pure spin-density-wave state evolves
into a state with coexisting spin- and charge-density waves. The
spin-wave velocity is reduced when several density waves coexist in the
ground state, and may even vanish at large $V$. The effect of
dimerization along the chain is also considered. 
\end{abstract}

\vspace{-3mm}
\pacs{PACS Numbers: 72.15.Nj, 75.30.Fv}
\vspace{-2mm}

\bcols

\section{Introduction}
\setcounter{equation}{0}

Organic conductors of the tetramethyltetraselenafulvalene (TMTSF) and 
tetramethyltetrathiafulvalene (TMTTF) salts family often
exhibit density-wave (DW) instability at low temperature. 
\cite{Jerome,Yamaji,Gruner94} 
Recent experiments have shown that a $2k_{\rm F}$ spin-density wave (SDW) may
coexist with a $4k_{\rm F}$ and/or a $2k_{\rm F}$ charge-density wave
(CDW). \cite{Pouget,Kagoshima}   
(The quantity  $k_{\rm F}$ denotes the one-dimensional Fermi wave vector and   
2$k_{\rm F}$ is the nesting wave vector for the SDW.) Furthermore, these
CDW's seem to be of pure electronic origin without any (significant)
contribution from the lattice. 
  
This unusual ground-state can be understood on the basis of a
mean-field theory for a quarter-filled one-dimensional system
in the presence of several kinds of Coulomb interaction.
Within an extended Hubbard model with on-site ($U$) and nearest-neighbor
($V$) interactions, it has been shown that a $4k_{\rm F}$ CDW may coexist
with the $2k_{\rm F}$ SDW when $V$ is strong enough. \cite{Seo} When the
next-nearest-neighbor interaction ($V_2$) is also taken into account,
three different ground states can be stabilized:
\cite{Kobayashi,TomioJPSJ,TomioCREST} (i) a pure $2k_{\rm F}$ SDW at
small $V$ and $V_2$, (ii) coexisting $2k_{\rm F}$ SDW and $4k_{\rm F}$ CDW at
large $V$, (iii) coexisting $2k_{\rm F}$ SDW, $2k_{\rm F}$ CDW and
$4k_{\rm F}$ SDW at large $V_2$. 
Although the SDW instability is driven by the on-site
repulsive interaction $U$, the nearest and next-nearest neighbor  
 interactions play a crucial role for the appearance of CDW's.
 
Following the standard analysis, 
\cite{Lee74,Takada84,Psaltakis84,Maki87,Maki90a,Maki90b,Suzumura90,Brazovskii76}
fluctuations around the mean-field ground-state have been studied. For
a quarter-filled system, commensurability effects with the underlying
crystal lattice pin the DW's and produce a gap in the sliding
modes. \cite{Suzumura_JPSJ95} Surprisingly, this gap vanishes at the
boundary between the pure $2k_{\rm F}$ SDW and the coexisting 
2$k_{\rm F}$ SDW and 4$k_{\rm F}$ CDW. 
\cite{Suzumura97} The spin-wave modes have been studied only
within the Hubbard model ($V=V_2=0$).\cite{TanemuraPTP} The spin-wave
velocity decreases monotonically with increasing $U$, in qualitative
agreement with the exact solution of the one-dimensional Hubbard
model. \cite{Schulz_int}  

In this paper, we study the spin-wave modes in presence of the nearest and 
 next-nearest neighbor 
 interactions ($V,V_2\neq 0$). We consider a one-dimensional
system, assuming that long-range order is stabilized by (weak)
interchain coupling. Our analysis is based on a functional integral  
formulation \cite{Wen88,Schulz90,Ting91,Dupuis} which allows a simple
treatment of the spin-wave modes even in the presence of these 
 interactions. The electron-electron interaction is treated
within (Hartree-Fock) mean-field theory, while the SU(2) spin rotation
symmetry is maintained by introducing a fluctuating spin-quantization
axis in the functional integral. Transverse spin-wave modes then
correspond to fluctuations of the spin-quantization axis around its
mean-field value. 

In Secs.~II and III, we extend the derivation of Ref.~\onlinecite{Dupuis} from the
incommensurate to the commensurate case. Previous mean-field
results\cite{TomioJPSJ} are recovered within a saddle point
approximation. Then we derive the effective action of the spin-wave
modes and obtain the spin-wave velocity. In Sec.~IV, the spin-wave
velocity is calculated as a function of $V,V_2$ and the dimerization
along the chain. Section V is devoted to discussion.

\section{Path integral formulation}

We consider a one-dimensional electron system at quarter-filling
with dimerization along the chain. Within the extended Hubbard model,
the Hamiltonian is given by  
\begin{eqnarray} 
 \label{H}
	H   &=&  H_0 + H_I  \;\;,  \\
 \label{H0}
        H_0 
	    &=& - \sum_{\sigma, n, n'}
	 t_{n n'} \psi_{n\sigma}^\dagger  \psi_{n'\sigma} \;\;, \\
 \label{HI}
        H_I &=&  U \sum_{n} n_{n \uparrow} n_{n \downarrow}
          +  V \sum_{n} 
           (\psi_{n}^{\dagger} \psi_{n})
                 (\psi_{n+1}^{\dagger} \psi_{n+1}) 
\nonumber \\ & & {}
          +  V_2 \sum_{n} 
           (\psi_{n}^{\dagger} \psi_{n})
                 (\psi_{n+2}^{\dagger} \psi_{n+2})
                 \nonumber \\ 
            &=&  {}
	      - \frac{U}{4} \sum_{n} 
              (\psi_n^{\dagger} \sigma_z \psi_n)^2 
	      + \sum_{n, n'} 
              (\psi_n^{\dagger} \psi_n) 
                V_{n n'} 
              (\psi_{n'}^{\dagger} \psi_{n'}) \;\;, 
\nonumber \\
\end{eqnarray}
where $\psi_{n}=(\psi_{n \uparrow},\, \psi_{n \downarrow})^{t}$,
$n_{n \sigma} = \psi_{n \sigma}^\dagger \psi_{n \sigma}$, and $\psi_{n
\sigma}^{\dagger}$ is the creation  operator of an electron with spin
$\sigma (=\uparrow, \downarrow)$ at the lattice site $n$. The
transfer  integral in the kinetic term $H_0$ is defined by
\begin{eqnarray}
 \label{transfer}
t_{n n'} &=& \left \lbrace 
\begin{array}{ll}
t-(-1)^n t_{\rm d} & {\rm for} \,\,\, n'=n+1 \;\;, \\
t+(-1)^n t_{\rm d} & {\rm for} \,\,\,  n'=n-1 \;\;, \\
0 & {\rm otherwise} \;\;, 
\end{array} \right.  
\end{eqnarray}
where a finite $t_{\rm d}$ is due to the dimerization. The interaction
Hamiltonian is expressed in terms of the Hubbard interaction $U$ and
the density-density interaction $V_{nn'}$ defined by
\begin{eqnarray}
 \label{interaction}
V_{n n'} &=& \left \lbrace 
\begin{array}{ll}
U/4   & {\rm for} \,\,\, n'=n \;\;, \\
V/2   & {\rm for} \,\,\, n'=n \pm 1 \;\;, \\
V_2/2   & {\rm for} \,\,\, n'=n \pm 2 \;\;, \\
0 & {\rm otherwise} \;\;, 
\end{array} \right.
\end{eqnarray}
where $V$ ($V_2$) is the coupling constant for nearest
(next-nearest) neighbor-site interaction ($U,V,V_2\geq 0$). 

In order to derive the effective action for the spin-wave modes,
we write the partition function $Z$ as a path integral:  
\begin{eqnarray}
            \label{Zstart}
 Z &=& \int {\cal D}\psi^{\dagger}{\cal D}\psi ~e^{-{\cal S}[\psi^{\dagger}, \psi]} \;\;,
                          \\
 \label{S}
 {\cal S} & = & \int d\tau \left[ \sum_{n} \psi_{n}^{\dagger} 
      \left(\partial_{\tau} - \mu \right) \psi_{n} 
    + H [\psi^{\dagger}, \psi] \right] \;\;,
\end{eqnarray}
where the action ${\cal S}$ is a function of the Grassmann variable
$\psi$. $\tau$ is a Matsubara time-varying between $0$ and
$1/T$. Following Refs.~\onlinecite{Schulz90} and \onlinecite{Dupuis}, we now introduce the
new field $\phi$ defined by 
\begin{eqnarray} \label{defphi}
 \psi_{n} &=& R_{n} \phi_{n} \;\;, \nonumber \\
 R_{n} \sigma_{z} R_{n}^{\dagger} &=& 
   \mbox{\boldmath $\sigma$} \cdot {\bf n}_{n} \;\;, 
\end{eqnarray}
where $R_{n}$ is an SU(2)/U(1) unitary matrix and ${\bf n}_n$ is a 
unit vector which gives the direction of the spin-quantization axis at
site $n$ and time $\tau$ for the field $\phi$. 
Substituting Eq.~(\ref{defphi}) into Eq.~(\ref{S}),  
the action is rewritten as  ${\cal S} = {\cal S}_0 + {\cal S}_I$, where 
\begin{eqnarray}
  \label{S0-2}
 {\cal S}_0 
      &=& \int d\tau \left\{ \sum_{n} \phi_{n}^{\dagger} 
(\partial_{\tau} - \mu + R_{n}^{\dagger} \partial_{\tau} R_{n}) \phi_{n}
\right.  \nonumber \\
  & & 
\left.
	 - \sum_{n,n'} \phi_{n}^{\dagger} R_{n}^{\dagger} t_{nn'}
	 R_{n'} \phi_{n'} \right\}  
                        \;\;, \\
  \label{SI-2}
 {\cal S}_I &=& \int d\tau \left\{ - \frac{U}{4} \sum_{n} \rho_{sn}^2 
	  + \sum_{n,n'} \rho_{cn} V_{nn'} \rho_{cn'} \right\} \;\,.
\end{eqnarray}
$\rho_{cn} = \phi_{n}^{\dagger} \phi_{n}$ 
and $\rho_{sn} = \phi_{n}^{\dagger} \sigma_{z} \phi_{n}$
are the charge- and spin-density operators.
The quantities $\sigma_x, \sigma_y$ and $\sigma_z$ are Pauli matrices.  
Note that ${\cal S}_I$ is invariant under the transformation $\psi\to\phi$,
since the interaction is invariant with respect to 
spin rotations. It is convenient to rewrite the action as 
\begin{eqnarray} \label{S3}
 {\cal S} &=& \int d\tau \left\{ \sum_{n} \phi_{n}^{\dagger} 
          (\partial_{\tau} - \mu - A_{0n}) \phi_{n} 
 \right. \nonumber \\
 & & 
 \left.
	  - \sum_{n,n'} \phi_{n}^{\dagger}  
	  t_{nn'} \exp\!\left(-i \int_{n}^{n'} dl A_{xl} \right) \phi_{n'} 
\right.
\nonumber \\
  & & \left. {}
 -  \frac{U}{4} \sum_{n} \rho_{sn}^2 
	 + \sum_{n,n'} \rho_{cn} V_{nn'} \rho_{cn'} \right\} \;\,.
\end{eqnarray}
where the SU(2) gauge fields $A_0$ and $A_x$ are defined by  
{\setcounter{enumi}{\value{equation}}
\addtocounter{enumi}{1}
\setcounter{equation}{0} 
\renewcommand{\theequation}{\arabic{section}.\theenumi\alph{equation}}
\begin{eqnarray} 
  \label{A0}
 A_{0n} & \equiv & - R^{\dagger}_n \partial_{\tau} R_n  \;, \\
  \label{Ax}
 \exp\!\left( -i \int_{n}^{n+\delta} dl A_{xl}\right) & \equiv & 
                R_{n}^{\dagger}  R_{n+\delta} \;, 
\hspace{0.2cm} ( \delta = \pm 1 )  \;\,.
\end{eqnarray}
\setcounter{equation}{\value{enumi}}}%
The lattice spacing is taken as  unity. Using the
Stratonovich-Hubbard identity\cite{Amit}, the interaction part of the
action is rewritten as (note that $U, V, V_2 > 0$) 
\bleq
\begin{eqnarray}
\label{deltac}
 \exp\!\left( - \sum_{n,n'} \int d\tau \rho_{cn} V_{nn'} \rho_{cn'} 
\right) 
 &=& \int {\cal D} \Delta_c  
  \exp\!\left(- \sum_{n,n'}\int d\tau \Delta_{cn} V_{nn'}^{-1} \Delta_{cn'} 
 + ~2i \sum_{n}\int d\tau \Delta_{cn} \rho_{cn} \right)  \;\;, \\
\label{deltas}
 \exp\!\left( \frac{U}{4} \sum_{n} \int d\tau \rho_{sn}^2 \right)
 &=& \int {\cal D} \Delta_{s} 
  \exp\!\left(-\frac{1}{U}\sum_{n}\int d\tau \Delta_{sn}^2 
 +  ~\sum_{n}\int d\tau \Delta_{sn} \rho_{sn} \right)  \;\;,
\end{eqnarray}
\eleq%
where   
 $\Delta_{cn}$ and $\Delta_{sn}$ are (real) auxiliary fields.  
By using Eqs.~(\ref{deltac}) and (\ref{deltas}), 
 the final form of the partition function is given by   
\begin{eqnarray}
\label{Z}
 Z &=& \int {\cal D} \Delta_{c} {\cal D} \Delta_{s} \int {\cal D} {\bf n} 
     \int {\cal D}\phi^{\dagger}{\cal D}\phi ~e^{-({\cal S}_0 + {\cal S}_I)} \;\;,  \\
\label{S04}
 {\cal S}_0 &=& \int d\tau \left\{ \sum_{n} \phi_{n}^{\dagger} 
          (\partial_{\tau} - \mu - A_{0n}) \phi_{n} 
\right. \nonumber \\
 & & \left. {}
          - \sum_{n,n'} \phi_{n}^{\dagger}  
          t_{nn'} \exp\left(-i \int_{n}^{n'} dl A_{xl} \right)
\phi_{n'} \right\} 
\;\;, \\
\label{SI4}
 {\cal S}_I &=& \int d\tau \left\{ \sum_{n} 
          \left[ \frac{1}{U} \Delta_{sn}^2  
	  - \Delta_{sn} \rho_{sn} 
          - 2 i \Delta_{cn} \rho_{cn} \right]
\right. \nonumber \\
 & & \left. {}
          + \sum_{n,n'} \Delta_{cn} V_{nn'}^{-1} \Delta_{cn'} \right\} 
\;\;,
\end{eqnarray}
where $V_{nn'}^{-1}=V_{n'n}^{-1}$. 

\section{Effective Action for the Spin-Wave Mode} 
 In this section, we derive  the action corresponding to the spin-wave 
  modes at quarter-filling. First, we reproduce the mean-field result
  of Ref.~\onlinecite{TomioJPSJ} within a saddle-point approximation. Then we
  consider transverse spin fluctuations arising from the dynamics of
  the spin-quantization axis.

\setcounter{equation}{0}
\subsection{Mean-field solution}

The standard mean-field solution is recovered from a saddle-point
 approximation with ${\bf n}=\hat{z}$ at each lattice site. One then has
 $R_{n}=1$ and $A_0=A_x=0$.  

 By minimizing the free energy with respect to $\Delta_{sn}$ and
 $\Delta_{cn}$, we obtain the self-consistent mean-field equations  
\begin{eqnarray}
\label{deltas2}
 \Delta_{sn} &=& \frac{U}{2} \left< \rho_{sn} \right>_{\rm MF} \;\;,  
\\
\label{deltac2}
 \Delta_{cn} &=& i \sum_{n'} V_{nn'} \left< \rho_{cn'} \right>_{\rm
 MF} \;\,.
\end{eqnarray}
 The average $\left< \;\; \right>_{\rm MF}$ is to be calculated with 
the mean-field action    
\begin{eqnarray} \label{SMF}
 {\cal S}_{\rm MF} &=& \beta \sum_{n} \frac{1}{U} \Delta_{sn}^2 
              + \beta \sum_{n,n'} \Delta_{cn} V_{nn'}^{-1} \Delta_{cn'} 
\nonumber \\
            & &  + {} 
  \int d\tau \left\{ \sum_{n} \phi_{n}^{\dagger} 
         \left( \partial_{\tau} -\mu - 2 i \Delta_{cn}  
         - \Delta_{sn} \sigma_{z} \right) \phi_{n}
\right. \nonumber \\
 & & \left. {}
         - \sum_{n,n'} \phi_{n}^{\dagger} t_{nn'} \phi_{n'} \right\} \;\,.
\end{eqnarray}

At quarter-filling, the mean-fields $\left<\rho_{sn}\right>_{\rm MF}$  and
$\left<\rho_{cn}\right>_{\rm MF}$ are periodic with a periodicity of four
lattice spacings. They can be written as 
\begin{eqnarray}
\label{Sm}
 \left< \rho_{sn} \right>_{\rm MF} &=& 
     \sum_{m=0}^{3} S_{m} \,e^{i mQ_0 n} 
                     \;\;,  \\
\label{Dm}
 \left< \rho_{cn} \right>_{\rm MF} &=& 
     \sum_{m=0}^{3} D_{m} \,e^{i mQ_0 n} \;\;, 
\end{eqnarray}
where $Q_0=2k_{\rm F}=\pi/2$. 
Since $\left< \rho_{cn} \right>_{\rm MF}$ and 
 $\left< \rho_{sn} \right>_{\rm MF}$ are  real quantities, one finds 
$D_0=D_0^*,  D_1=D_3^*, D_2=D_2^*$ and  
$S_0=S_0^*,  S_1=S_3^*, S_2=S_2^*$. 
In Eqs.~(\ref{Sm}) and (\ref{Dm}),   
 $S_0 =0$  due to the absence of ferromagnetism 
 and  $D_0=1/2$ for a  quarter-filled band.    
 From Eqs.~(\ref{deltas2})-(\ref{Dm}), 
 the final form of the mean-field action is obtained as 
\bleq
\begin{eqnarray} \label{SMF2}
 {\cal S}_{\rm MF} &=& \beta N \left[ 
   -\frac{U}{16}-\frac{U}{2} (|D_1|^2-|S_1|^2)
   -\frac{U}{4} (D_2^2-S_2^2) -V (\frac{1}{4}-D_2^2) 
   -V_2 (\frac{1}{4}-2|D_1|^2+D_2^2) \right]
 \nonumber \\
 & & {} +  \int d\tau \left\{ \sum_{k} \phi_{k}^{\dagger} 
     \left( \partial_{\tau} - \mu + \frac{U}{4} + V + V_2 
     - 2t \cos k \right) \phi_{k}  \right. \nonumber \\
 & & {} + 
      \left[ \sum_{k} \phi_{k}^{\dagger} \left(
 \frac{U}{2} (D_1 -S_1 \sigma_{z}) -2 V_2 D_1 \right) \phi_{k-Q_0} 
     + c.c. \right]  \nonumber \\
 & & {} +  \left.
      \sum_{k} \phi_{k}^{\dagger} \left(
 \frac{U}{2} (D_2 -S_2 \sigma_{z}) - 2V D_2 + 2V_2 D_2 -2 i t_{\rm d} \sin k
   \right) \phi_{k-2Q_0} \right\} \;\;,
\end{eqnarray}
\eleq
where 
$
 \phi_{k} = (1/\sqrt{N}) \sum_{n} \,e^{ - i k n} \phi_{n} 
$
and  $N$ is the number of lattice sites. 
The action (\ref{SMF2}) agrees with the mean-field Hamiltonian 
obtained previously by the conventional method. \cite{TomioJPSJ}

\subsection{Fluctuations}

In the long-wavelength limit, collective modes can be separated into
sliding (charge) modes and spin-wave modes. In this paper, we consider
only transverse (acoustic) spin-wave modes (i.e. magnons). These modes
show up in the fluctuations of the unit vector field ${\bf n}$. They
do not couple to charge modes and gapped spin-wave modes. We shall make the
following two approximations: (i) We neglect the coupling to
long-wavelength spin fluctuations [$\Delta_{s}(q)$ with $|q| \ll
Q_0$]. In the Hubbard model ($V=V_2=0$), this coupling is known to
renormalize the spin-wave velocity by the factor $[1-UN(0)]^{1/2}$
 in the weak-coupling limit\cite{Poilblanc87}
[$N(0)$ is the density of states at the Fermi
level]; (ii) We also neglect any possible coupling to spin
fluctuations at wave-vector $2Q_0+q$ [$\Delta_{s}(2Q_0 + q)$ with
$|q| \ll Q_0$]. \cite{note}

When two SDW's coexist in the ground-state, our formalism can only
yield the ``in-phase'' modes where the two spin-density waves
oscillate in phase. It misses the modes where the oscillations are
out-of-phase. \cite{Dupuis1} These modes are gapped and do not couple
to the ``in-phase'' modes considered in this paper. 

Before proceeding with the spin-wave mode analysis, let us discuss
the limit of validity of our approach. The spin-wave modes will be
obtained by expanding about the (Hartree-Fock) mean-field state. Such
an approach should hold (at least qualitatively) as long as the
interaction is smaller than the 
bandwidth, i.e. $U,V,V_2 \lesssim 4t$. Nevertheless, it does not
necessary break down in the strong-coupling limit. In the context of
the two-dimensional Hubbard model, Schrieffer {\it et al.} have shown
that an RPA analysis of the fluctuations about the mean-field state in
the limit $U\gg t$ agrees with the conclusions obtained from the
Heisenberg model with exchange constant
$J=4t^2/U$. \cite{Schrieffer89} 

Another limitation of our approach comes from the analysis of the
fluctuations of the unit vector ${\bf n}$. As will become clear below,
the main assumption is that ${\bf n}$ is a slowly varying field, thus
allowing a gradient expansion. Whereas this assumption is perfectly
valid in the weak-coupling limit ($U,V,V_2\lesssim 4t$), it breaks
down in the strong-coupling limit. In the latter, one should write
${\bf n}_n={\bf n}^{\rm slow}_n+\cos(n\pi/2){\bf L}_n$ where
${\bf n}^{\rm slow}_n$ is a slowly varying field and ${\bf
L}_n$ a small perpendicular component (${\bf L}_n\cdot
{\bf n}^{\rm slow}_n=0$ and $|{\bf L}_n|\ll |{\bf n}^{\rm
slow}_n|\simeq 1$). \cite{Auerbach,Schulz90} The effective
action of the spin-wave modes, 
$S_{\rm eff}[{\bf n}^{\rm slow}]$, is then obtained by integrating out
both the fermions and the (small) transverse component ${\bf L}_n$. 
For $V=V_2=0$, this allows to interpolate
smoothly between the weak-coupling regime and the strong-coupling
regime which is well described by the Heisenberg model. \cite{Schulz90}

Long-wavelength transverse spin fluctuations correspond to 
fluctuations of the SU(2) gauge fields $A_0$ and $A_x$
 [Eqs.~(\ref{A0}) and (\ref{Ax})] which are rewritten as 
\begin{eqnarray}
   \label{A00}
 A_{0n} &=& \sum_{\nu=x,y,z} A_{0n}^{\nu} \sigma_{\nu}  \;\;, \\
       \label{Axx}
 A_{xn} &=& \sum_{\nu=x,y,z} A_{xn}^{\nu} \sigma_{\nu}  \;\,. 
\end{eqnarray}
 From eqs.~(\ref{S3}), (\ref{SMF2}), (\ref{A00}) and (\ref{Axx}), we
 write the action of the spin degrees of freedom as 
\begin{eqnarray}
 {\cal S} &=&  {\cal S}_{\rm MF}  - \sum_{n} \int d\tau \phi_{n}^{\dagger}
      A_{0n} \phi_{n} 
 \nonumber \\
 & & {}
\hspace{-3mm}
 - \sum_{n,n'} \int d\tau \phi_{n}^{\dagger} 
 \left[ t_{nn'} \exp\!\left(-i \int_{n}^{n'} dl A_{xl} \right)-t_{nn'} 
\right] \phi_{n'}
 \;\,. \nonumber \\
\end{eqnarray}
To order  $O$($A_x^2$) we obtain 
\bleq
\begin{eqnarray}
 {\cal S} & = & {\cal S}_{\rm MF}  - \sum_{n} \int d\tau \phi_{n}^{\dagger}
      A_{0n} \phi_{n} 
 - \sum_{n,n'} \int d\tau  t_{nn'} \phi_{n}^{\dagger} 
 \left(  
 -\frac{i}{2}(n-n') \left(A_{xn}+A_{xn'} \right) -\frac{1}{2} A_{xn}^2 \right) \phi_{n'} 
         \nonumber \\
          \label{Sfluctuation}
  & = & {\cal S}_{\rm MF} - \sum_{n} \sum_{\mu=0,x} \sum_{\nu=x,y,z}
      \int d\tau~ j_{\mu n}^{\nu} A_{\mu n}^{\nu} 
    + {\cal S}_x^{\rm dia}  \;\;,
\end{eqnarray}
\eleq
where $j_{xn}^{\nu}$, $j_{0n}^{\nu}$ and  ${\cal S}_x^{\rm dia}$ 
 are given by 
\begin{eqnarray}
j^\nu_{xn} &=& - \frac{i}{2} \sum_{\delta=\pm 1} \delta \left[
t_{n,n+\delta} \phi^\dagger_n \sigma_\nu \phi_{n+\delta} 
+ t_{n-\delta,n} \phi^\dagger_{n-\delta} \sigma_\nu \phi_n \right] 
\;\;, 
\nonumber \\
           \label{Pi235} \\
  j_{0 n}^{\nu} &=& \phi_n^\dagger \sigma_{\nu} \phi_n 
                        \;\;,           \\ 
 {\cal S}_x^{\rm dia} &=&  \frac{1}{2} \sum_{n,n'} \sum_{\nu,\nu'} t_{nn'}
     \int d\tau~ \phi_{n}^{\dagger} \sigma_{\nu} \sigma_{\nu'} \phi_{n'}
     A_{xn}^\nu A_{xn}^{\nu'} \;\,. 
\end{eqnarray}
The second term  of Eq.~(\ref{Sfluctuation})
 denotes the coupling of the gauge field $A_{\mu n}^\nu$ to the 
spin current ($j_{xn}^{\nu}$) and spin density ($j_{0n}^{\nu}$). The
 last term of Eq.~(\ref{Sfluctuation}), ${\cal S}_x^{\rm dia}$, is the
 diamagnetic contribution. \cite{Dupuis}    

The effective action of the gauge field is obtained by 
integrating out the fermions in the partition function.
By substituting Eq.~(\ref{Sfluctuation}) into Eq.~(\ref{Z}),
one obtains the effective action up to $O(A^2)$ as  
\begin{eqnarray}
\label{SeffA}
 {\cal S}_{\rm eff}[A_{\mu}^{\nu}] &=&  \left< {\cal S}_{x}^{\rm dia}
\right>_{\rm MF}  - \sum_{n,\mu,\nu} 
   \int d\tau \left< j_{\mu n}^{\nu} \right>_{\rm MF} A_{\mu n}^{\nu} 
     \nonumber \\
 & & {} -
  \frac{1}{2} \sum_{n,n'} \sum_{\mu,\mu',\nu,\nu'} 
   \int d\tau d\tau' A_{\mu n}^{\nu}(\tau) 
     \nonumber \\
 & & {} \times 
      \Pi_{j_{\mu}^{\nu} j_{\mu'}^{\nu'}}(n,\tau,n',\tau') 
       A_{\mu'n'}^{\nu'}(\tau')  \;\;,   
\end{eqnarray}
where 
\begin{eqnarray}
\label{Pi245}
  \Pi_{j_{\mu}^{\nu} j_{\mu'}^{\nu'}}(n,\tau,n',\tau')  =   
\left< j_{\mu n}^{\nu}(\tau) j_{\mu' n'}^{\nu'}(\tau') \right>_{\rm MF}
           \;\;,         \\
 \label{Sxdia}
\left< {\cal S}_{x}^{\rm dia} \right>_{\rm MF}  = 
\frac{1}{2} \sum_{n,n',\nu} t_{nn'} \int d\tau 
  \left< \phi_n^\dagger \phi_{n'} \right>_{\rm MF} (A_{xn}^{\nu})^2 \;\,.
\end{eqnarray}
The quantity $\Pi_{j_{\mu}^{\nu} j_{\mu'}^{\nu'}}$ denotes  
 the current-current correlation function in the mean-field state.  
 We note that 
$\left< j_{\mu n}^{\nu} \right>_{\rm MF}=0$
in the long-wavelength limit \cite{Berry}
and that 
 $A_{\mu n}^{\nu}$ is  of the order $O(\nabla)$.
To order $O(\nabla^2)$), we obtain
\begin{eqnarray}
\label{Seff251}
 {\cal S}_{\rm eff} &=& -\frac{1}{2} \sum_{\tilde{q}} 
\left\{
  \left< K \right>_{\rm MF} \sum_{\nu=x,y,z} |A_{x}^{\nu}(\tilde{q})|^2 
\right. \nonumber \\
 & & \left. {}
\hspace{-4mm}
 + \sum_{\mu,\mu'(=0,x)} 
   \sum_{\nu,\nu'} 
   A_{\mu}^{\nu}(\tilde{q}) A_{\mu'}^{\nu'}(-\tilde{q}) 
   ~\Pi_{j_{\mu}^{\nu} j_{\mu'}^{\nu'}}(\tilde{q})  
\right\}  \;,
\end{eqnarray}
\begin{eqnarray}
 \left< K \right>_{\rm MF} = \left< 
 -\frac{1}{N} \sum_{n,n'} t_{nn'} \phi_n^\dagger \phi_{n'} \right>_{\rm MF}
\;\;,
\end{eqnarray}
where $\left< K \right>_{\rm MF}$ is the mean value of the kinetic
energy per site in the mean-field state. $\tilde{q}=(q, ~i \Omega)$ and
$\Omega$ is a bosonic Matsubara frequency. 
The quantity $\Pi_{j_{\mu}^{\nu} j_{\mu'}^{\nu'}}(\tilde{q})$ 
  is the Fourier transform of Eq.~(\ref{Pi245}) 
 with respect to $n$  and $\tau$. In Eq.~(\ref{Seff251}), it can be
 evaluated at $\tilde{q} = 0$ since $A_{\mu}^{\nu} \propto O(\nabla)$. 
Note that 
$
 \Pi_{j_{\mu}^{x} j_{\mu'}^{y}} = 
 \Pi_{j_{\mu}^{x} j_{\mu'}^{z}} =
 \Pi_{j_{\mu}^{y} j_{\mu'}^{z}} = 0
$
and  
$
 \Pi_{j_{0}^{\nu} j_{x}^{\nu}} (\tilde{q}) ~\big|_{\tilde{q}=0} = 0
$. 
 Taking the continuum limit $n \rightarrow \xi$ (with $\xi$ a real
 continuous variable) and writing 
$A^{\nu}_{\mu n}=A^{\nu}_{\mu}(\xi,\tau)$, the effective action 
(\ref{Seff251}) is rewritten as 
\bleq
\begin{eqnarray}
  \label{Seff256}
  {\cal S}_{\rm eff} &=& -\frac{1}{2} \sum_{\tilde{q}} \sum_{\nu=x,y,z} 
\left\{
  \left< K \right>_{\rm MF} |A_{x}^{\nu}(\tilde{q})|^2 
 + \sum_{\mu} 
   |A_{\mu}^{\nu}(\tilde{q})|^2   
   ~\Pi_{j_{\mu}^{\nu} j_{\mu}^{\nu}} \right\}  
                     \nonumber \\
  &=& -\frac{1}{2} \int d\xi d\tau 
\left\{
 \left( \left< K \right>_{\rm MF} + \Pi_{j_{x}^{x} j_{x}^{x}} \right) 
 \sum_{\nu=x,y} A_{x}^{\nu \, 2}(\xi,\tau)
 + 
\Pi_{j_{0}^{x} j_{0}^{x}} 
\sum_{\nu=x,y} A_{0}^{\nu \, 2}(\xi,\tau) \right\}   
                             \nonumber \\
 & & {} - 
  \frac{1}{2} \sum_{\tilde{q}} 
\left\{
 \left( \left< K \right>_{\rm MF} + \Pi_{j_{x}^{z} j_{x}^{z}} \right) 
 |A_{x}^{z}(\tilde{q})|^2 
 + \Pi_{j_{0}^{z} j_{0}^{z}}|A_{0}^{z}(\tilde{q})|^2 
\right\}  \;\;,
\end{eqnarray}
\eleq
where 
$
\Pi_{j_{\mu}^{\nu} j_{\mu}^{\nu}} \equiv 
\Pi_{j_{\mu}^{\nu} j_{\mu}^{\nu}}(\tilde{q}=0)
$
 and 
$ 
 \Pi_{j_{\mu}^{x} j_{\mu}^{x}} =  \Pi_{j_{\mu}^{y} j_{\mu}^{y}} 
$. 
Here we note the identities  
$
\left< K \right>_{\rm MF} + \Pi_{j_{x}^{z} j_{x}^{z}} = 0  
$ and  
$
 \Pi_{j_{0}^{z} j_{0}^{z}} = 0
$,
 which can be  deduced from the gauge invariance 
  of Eq.~(\ref{defphi}) (Appendix A). 
We have verified  numerically the validity of these
identities. Finally, noting that 
\cite{Schakel} 
\begin{eqnarray}
\sum_{\nu=x,y} A_{x}^{\nu \, 2}(\xi,\tau) & = &  
                 \frac{1}{4} (\partial_{\xi} {\bf n} )^2 \;\;, \\
 \sum_{\nu=x,y} A_{0}^{\nu \, 2}(\xi,\tau) & = & 
             - \frac{1}{4} (\partial_{\tau} {\bf n} )^2  \;\;,
\end{eqnarray}
we obtain the following final expression for the effective action of
the spin-wave modes \cite{Dupuis,Auerbach} (Appendix A): 
\begin{eqnarray}
   \label{Seff273}
 {\cal S}_{\rm eff} &=& \frac{1}{2} \int d\xi d\tau 
\left\{
 \chi (\partial_{\tau} {\bf n} )^2 
 + \rho (\partial_{\xi} {\bf n})^2 
\right\}  \;\;,
\end{eqnarray}
where $\chi$ and $\rho$ are the uniform transverse spin susceptibility and  
the spin stiffness, respectively:  
\begin{eqnarray}
   \label{chi274}
 \chi &=& \left< S_{\nu} S_{\nu} \right>_{\tilde{q}=0}^{\rm MF} 
 = \frac{1}{4} \Pi_{j_0^\nu j_0^\nu}  \;\;, 
\hspace{0.5cm} (\nu=x,y) \;\;, \\
               \label{rho275}
 \rho &=& - \frac{1}{4} 
  \left( 
\left< K \right>_{\rm MF} + \Pi_{j_x^{\nu} j_x^{\nu}} 
  \right)  \;\;, 
\hspace{0.5cm} (\nu=x,y) \;\,.
\end{eqnarray}
 From Eq.~(\ref{Seff273}) we deduce the spin-wave velocity
\begin{eqnarray}
  \label{spin_v}
 v = \left( \frac{\rho}{\chi} \right)^{\frac{1}{2}}  \;\,.
\end{eqnarray}
In the incommensurate case,
$ \Pi_{j_x^{\nu} j_x^{\nu}} \rightarrow  0$ 
in the weak coupling limit so that $\rho=-\left<K\right>_{\rm MF}/4$. 
\cite{Dupuis}
As shown in the next section, 
$\Pi_{j_x^{\nu} j_x^{\nu}}$ gives rise to a contribution of the same order 
 as $\left< K \right>_{\rm MF}$ in the quarter-filled case when   
  the on-site interaction $U$ is of the order of the bandwidth.    
In Eqs.~(\ref{chi274}) and (\ref{rho275}), 
$j_{\mu}^{\nu}$ and $K$ can be expressed as  
[$\phi_k=(\phi_{k\uparrow},\, \phi_{k\downarrow})^t$, $\nu=x,y$]
\begin{eqnarray}
  \label{j0}
j_0^{\nu}(\tilde{q}=0) &=& \frac{1}{\sqrt{N}} \sum_{k} 
             \phi_{k}^{\dagger} \sigma_{\nu} \phi_k  \;\;, \\
  \label{jx}
j_{x}^{\nu}(\tilde{q}=0) &=& \frac{1}{\sqrt{N}} \sum_{k} \left(
        2t \sin k \, \phi_{k}^{\dagger} \sigma_{\nu} \phi_k
\right. \nonumber \\
 & & \left. {}
      -2 i t_{\rm d} \cos k \, \phi_{k}^{\dagger} \sigma_{\nu} \phi_{k+2Q_0} 
           \right) \;\;, \\
  \label{Kft}
K &=& \frac{1}{N}\sum_{k} 
               \left( - 2t \cos k \, \phi_{k}^{\dagger} \phi_{k} 
\right. \nonumber \\
 & & \left. {}
      -2 i t_{\rm d} \sin k \, \phi_{k}^{\dagger} \phi_{k+2Q_0} 
               \right)  \;\,. 
\end{eqnarray}

\section{Spin-Wave Velocity}
\setcounter{equation}{0}

In this section, we evaluate the spin-wave velocity 
   at zero temperature ($T=0$). We take $t=1$ 
 and calculate  the velocity  normalized to its value at  
 $V=V_2=0$ and $t_{\rm d}=0$.

The phase diagram of the present model as a function of $V$ and $V_2$ 
 is shown in Fig.~1 for $U=4$ and $t_{\rm d}=0$ (solid curve).\cite{TomioJPSJ}
 For small $V$ and $V_2$, there is a
   pure 2$k_{\rm F}$ SDW state (region I). A large $V$ induces a phase with
 both a 2$k_{\rm F}$ SDW and 4$k_{\rm F}$ CDW (region II), while in the presence
 of a large $V_2$ there is coexistence between a 2$k_{\rm F}$ SDW, a
2$k_{\rm F}$ CDW and a 4$k_{\rm F}$ SDW (region III).  
 The dashed curve denotes  the boundary
  at which a first order transition occurs between II and III. 
The dash-dotted curve shows the phase diagram for $t_{\rm d}=0.1$.
The sliding modes are gapped in all three regions. However, the charge 
fluctuations become gapless at the transition
 between I and II. We discuss below the spin-wave velocity 
 [Eq.~(\ref{spin_v})] as a function of $V$ and $V_2$ for both $t_{\rm d}=0$
 and $t_{\rm d} \not= 0$.

\subsection{$U$ dependence ($V=V_2=0$ and $t_{\rm d}=0$)}

The spin stiffness $\rho$ and the susceptibility $\chi$ are shown in
  Fig.~2(a) as a function of $U$ for $V=V_2=0$ and $t_{\rm d}=0$.
Both  $\rho$ and $\chi$ are almost constant for small $U$  
  and decrease monotonically for large $U$. 
The inset shows  the corresponding 
  $U$-dependence for  $\left< K \right>_{\rm MF}$  and 
  $\Pi_{j_x^x j_x^x}$ which determine $\rho$ [Eq.~(\ref{rho275})].  
 A behavior similar to 
%
%
\begin{figure}[tb]
\begin{minipage}{8cm} 
\begin{center}
\vspace*{0mm}
\leavevmode
\epsfysize=7.4cm
   \epsffile{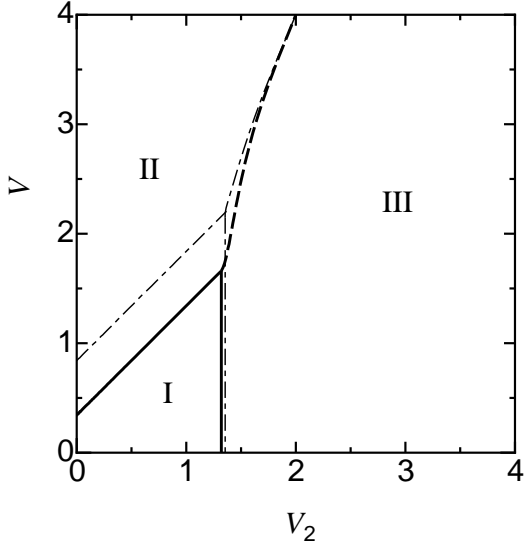}
\vspace*{3mm}
\caption[]{
 Phase diagram in the $V-V_2$ plane
for $U=4$ and $t_{\rm d}=0$ (Ref.~\onlinecite{TomioJPSJ}). 
The three different regions
correspond to: (I) pure 2$k_{\rm F}$ SDW, (II) coexisting  
2$k_{\rm F}$ SDW and 4$k_{\rm F}$ CDW, (III) coexisting
2$k_{\rm F}$ SDW, 2$k_{\rm F}$ CDW and 4$k_{\rm F}$ SDW. 
The dash-dotted curves denote the corresponding boundaries 
 for $t_{\rm d}=0.1$. 
}
\end{center}
\end{minipage}
\end{figure}
%
\begin{figure}[t]
\begin{minipage}{8cm} 
\begin{center}
\vspace*{0mm}
\leavevmode
\epsfysize=7.4cm
   \epsffile{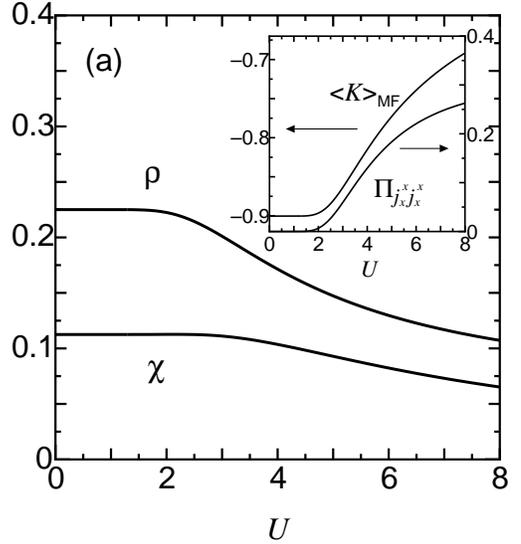}
\vspace*{5mm}
\leavevmode
\epsfysize=7.4cm
   \epsffile{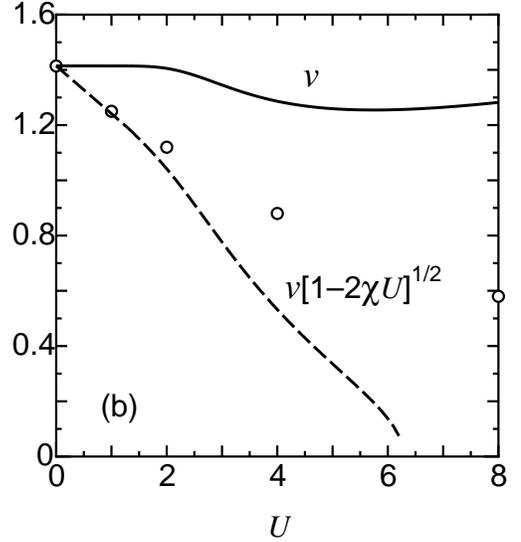}
\vspace*{3mm}
\caption[]{ (a)~
 $U$ dependence of $\rho$  and   $\chi$
    for $V=V_2=0$ and $t_{\rm d}=0$.  
 The inset shows  
 the $U$ dependence of $\left< K \right>_{\rm MF}$  and 
  $\Pi_{j_x^x j_x^x}$.
~(b)~
 $U$ dependence of $v$ (solid curve) and 
 $v[1- 2 \chi U]^{1/2}$ (dashed curve). The open circles denote the
exact result for the one-dimensional Hubbard model
(Ref.~\onlinecite{Schulz_int}).
}
\end{center}
\end{minipage}
\end{figure}
%
%
\noindent
the incommensurate case 
  is seen for $U \lesssim 2$:  
     $\Pi_{j_x^x j_x^x}$ is vanishingly small, and 
   $\chi$, $\rho$ and $\left< K \right>_{\rm MF}$ are almost constant 
   with respect to $U$. 
 The limiting values for small $U$ are given by 
    $\chi = 1/(2\sqrt{2}\pi) \simeq 0.113$,    
     $\left< K \right>_{\rm MF} = -2 \sqrt{2}/\pi \simeq -0.90$, and $v=\sqrt{2}$. 
 The variation of these quantities  for  $U \gtrsim 2$ 
 comes from  the effect of commensurability at quarter-filling. 
%
\\ \indent
In  Fig.~2(b) (solid curve), we show 
the spin-wave velocity $v$ [Eq.~(\ref{spin_v})], which 
 is almost independent of $U$ although slightly  
  suppressed at large $U$. 
 Here we note that we have neglected the coupling to long-wavelength
spin fluctuations. In the  Hubbard model ($V=V_2=0$), the  
spin-wave velocity $v=(\rho/\chi)^{1/2}$ becomes $(\rho/\chi)^{1/2}
[1- 2\chi U]^{1/2}$ when this coupling is taken into account within the
RPA. \cite{Poilblanc87} One obtains  $1-2\chi U = 1-UN(0)$
 in the weak-coupling limit 
 where  $N(0) =1/\sqrt{2} \pi$  at quarter-filling. 
In Fig.~2(b), we show $v$ and $v[1- 2\chi U]^{1/2}$ (dashed
curve).  The open circles denote the exact result for the
one-dimensional Hubbard model. \cite{Schulz_int} 
 For $U \lesssim 2$,  the RPA result  
turns out to be a good approximation, while the difference  becomes
 noticeable at larger $U$. 
 Nevertheless we use   $v=(\rho/\chi)^{1/2}$ 
 as a first step to examine the spin-wave velocity
 as the function of  $V$ and $V_2$.
The present calculation is performed  
 by choosing  $U=4$, 
  which leads to $v \simeq 1.29$ for  $V=V_2=0$ and $t_{\rm d}=0$. 

\subsection{$V$ dependence ($V_2=0$ and $t_{\rm d}=0$)}

Now we consider the $V$ dependence of the spin-wave velocity for
$V_2=0$, $t_{\rm d}=0$ and $U/t = 4$. Contrary to the weak-coupling limit
which can be studied analytically as the incommensurate
case,\cite{Dupuis} this intermediate coupling regime requires
numerical calculation.  
Figure 3 shows the $V$-dependence of $v$, $\chi$ and $\rho$ (all
quantities are normalized to their value at $V=V_2=0$ and $t_{\rm d}=0$). 
The arrow indicates the critical value $V_c=0.34$ separating regions I
($S_1 \not=0$) and II ($S_1,D_2 \not= 0$). In region II ($V>V_c$),
both $\rho$ and $\chi$ decrease for decreasing $V$. The stronger
decrease of $\rho$ results in a decrease of the spin-wave
velocity. For large $V$, both the spin stiffness
%
%
\begin{figure}[b]
\begin{minipage}{8cm}
\begin{center}
\leavevmode
\epsfysize=7.4cm
   \epsffile{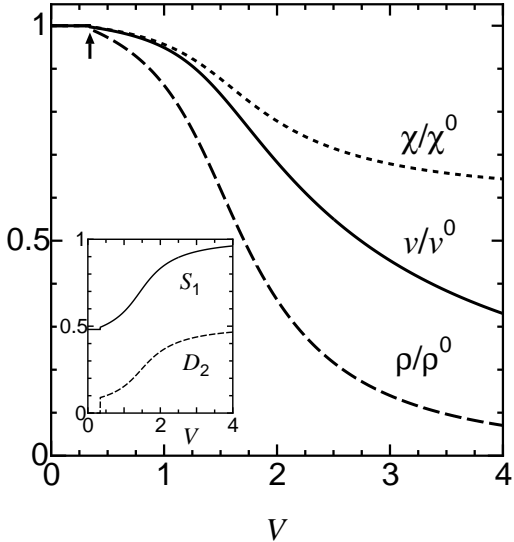}
\vspace*{2mm}
\caption[]{
 $V$ dependence of the spin-wave velocity velocity $v$, susceptibility $\chi$ 
and spin stiffness  $\rho$ for $U=4$, $t_{\rm d}=0$ and $V_2=0$. 
$v^0 (= 1.286)$, $\chi^0 (=0.103)$ and $\rho^0 (=0.171)$ 
 are the values for $V=V_2=0$.  
 There is a small jump at the critical value $V_c=0.34$,
 which is shown by the arrow. 
 The inset shows the $V$-dependence of the order parameters 
 $S_1$ (2$k_{\rm F}$ SDW) and $D_2$ (4$k_{\rm F}$ CDW).  
}
\end{center}
\end{minipage}
\end{figure}
\begin{figure}[t]
\begin{minipage}{8cm}
\begin{center}
\leavevmode
\epsfysize=7.4cm
   \epsffile{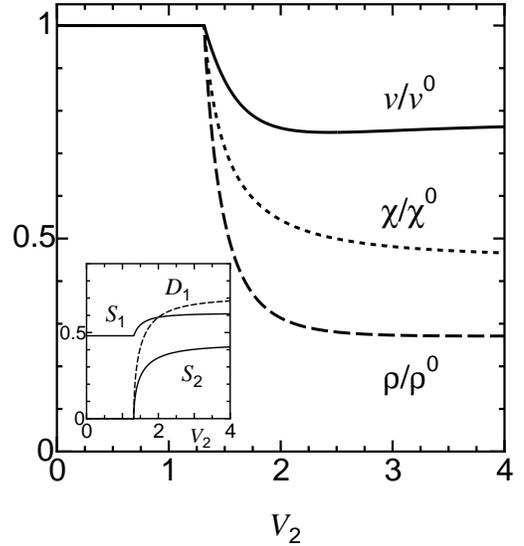}
\vspace*{2mm}
\caption[]{
 $V_2$ dependence of $v$, $\chi$ and $\rho$ 
    for $U=4$, $t_{\rm d}=0$ and $V=0$.  
 There is a cusp at the critical value  
   $V_{2c} = 1.32$  
    corresponding to the transition from state I to state III. 
 The inset shows the $V_2$-dependence of the order parameters 
 $S_1$, $D_1$ (2$k_{\rm F}$ CDW) and $S_2$ (4$k_{\rm F}$ SDW).   
}
\end{center}
\end{minipage}
\end{figure}
\vspace{3mm}
\noindent
 and the spin-wave
velocity vanish. It seems that the decrease of $v$ in region II
mainly comes from the reduction 
of kinetic energy due to the formation of the $4k_{\rm F}$-CDW. 
Note that the spin-wave velocity is discontinuous at the critical 
value $V=V_c$. The small jump at $V_c$ originates in the discontinuity
of $S_1$ and $D_2$ (see inset of Fig.~3) which is found only for $t_{\rm d} = 0$. 
\cite{Seo}

\subsection{$V_2$ dependence ($t_{\rm d}=0$)}

\vspace*{-3mm}
In this section, we analyze the $V_2$ dependence of the spin-wave
velocity for $U=4$, $t_{\rm d}=0$ and different values of $V$. 

Figure 4 shows $v/v^0$, $\chi/\chi^0$ and $\rho/\rho^0$ in the case
$V=0$ (the inset shows $S_1$, $D_1$ and $S_2$ as a function
of $V_2$). There is a transition between regions I and III at the critical
value $V_{2c}$. $v/v^0$, $\chi/\chi^0$ and $\rho/\rho^0$ are constant for
$V_2<V_{2c}$, and decrease for $V_2>V_{2c}$ (note that $v$ actually
slightly increases at large $V_2$). However, all these
quantities remain finite in the limit of large $V_2$. This is to be
contrasted to the large-$V$ limit (region II) where the spin-wave
velocity vanishes (Fig.~3). Such a behavior can be understood as
follows. For $V_2\to\infty$ (region III), the spin- and charge-density
waves in the ground-state are of the type ($\uparrow$,$\downarrow$,0,0) 
and (1,1,0,0), respectively. Our numerical calculation shows that this
behavior already shows up for $V_2/t\simeq 4$. In this limit ($V_2/t\gtrsim 4$), 
 the one-dimensional chain divides into independent two-site 
clusters. For this problem, one can find the exact expression of the
spin-wave velocity (Appendix B):
%
\begin{figure}[t]
\begin{minipage}{8cm} 
\begin{center}
\vspace*{0mm}
\leavevmode
\epsfysize=7.4cm
   \epsffile{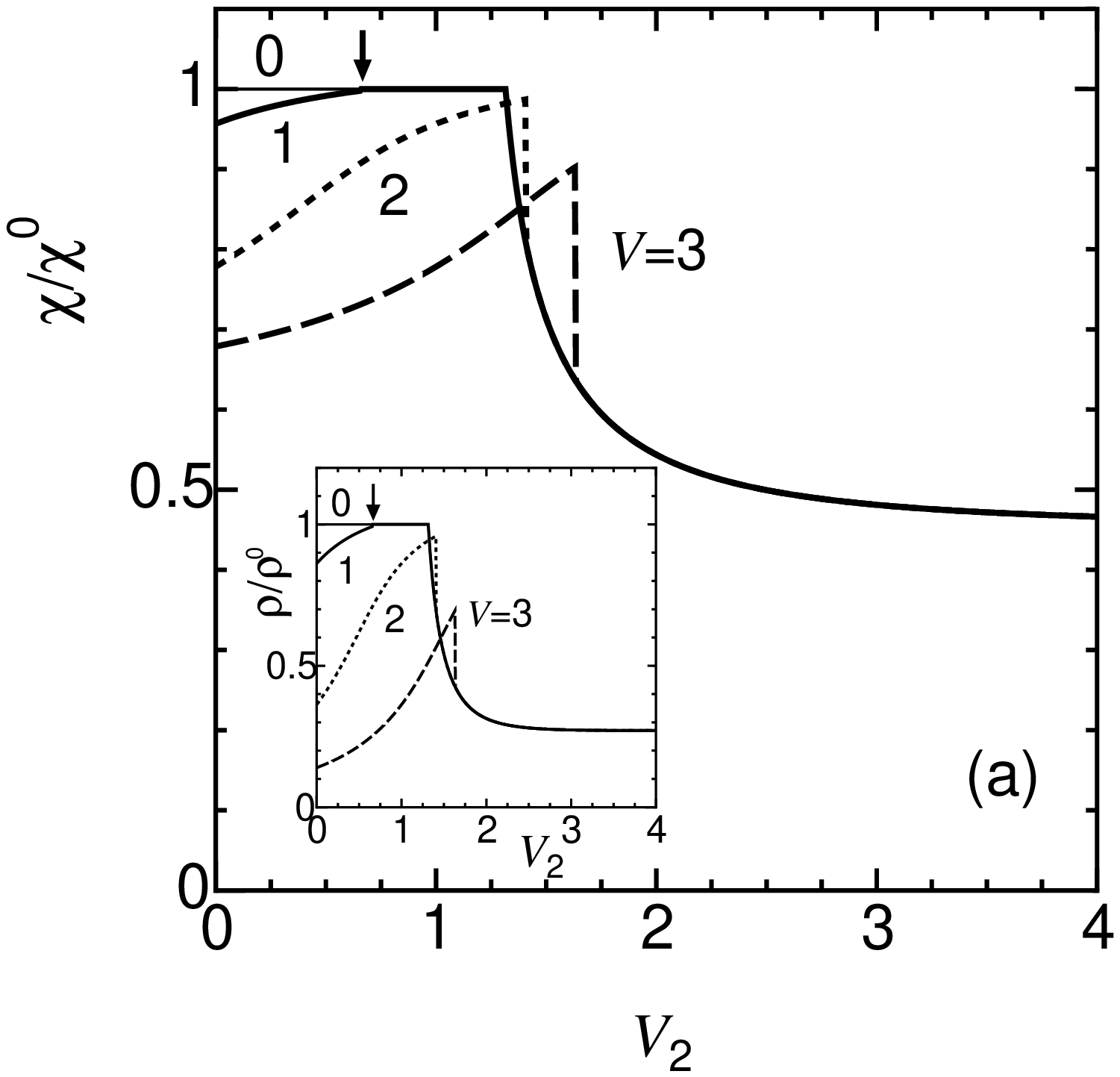}
\vspace*{3mm}
\leavevmode
\epsfysize=7.4cm
   \epsffile{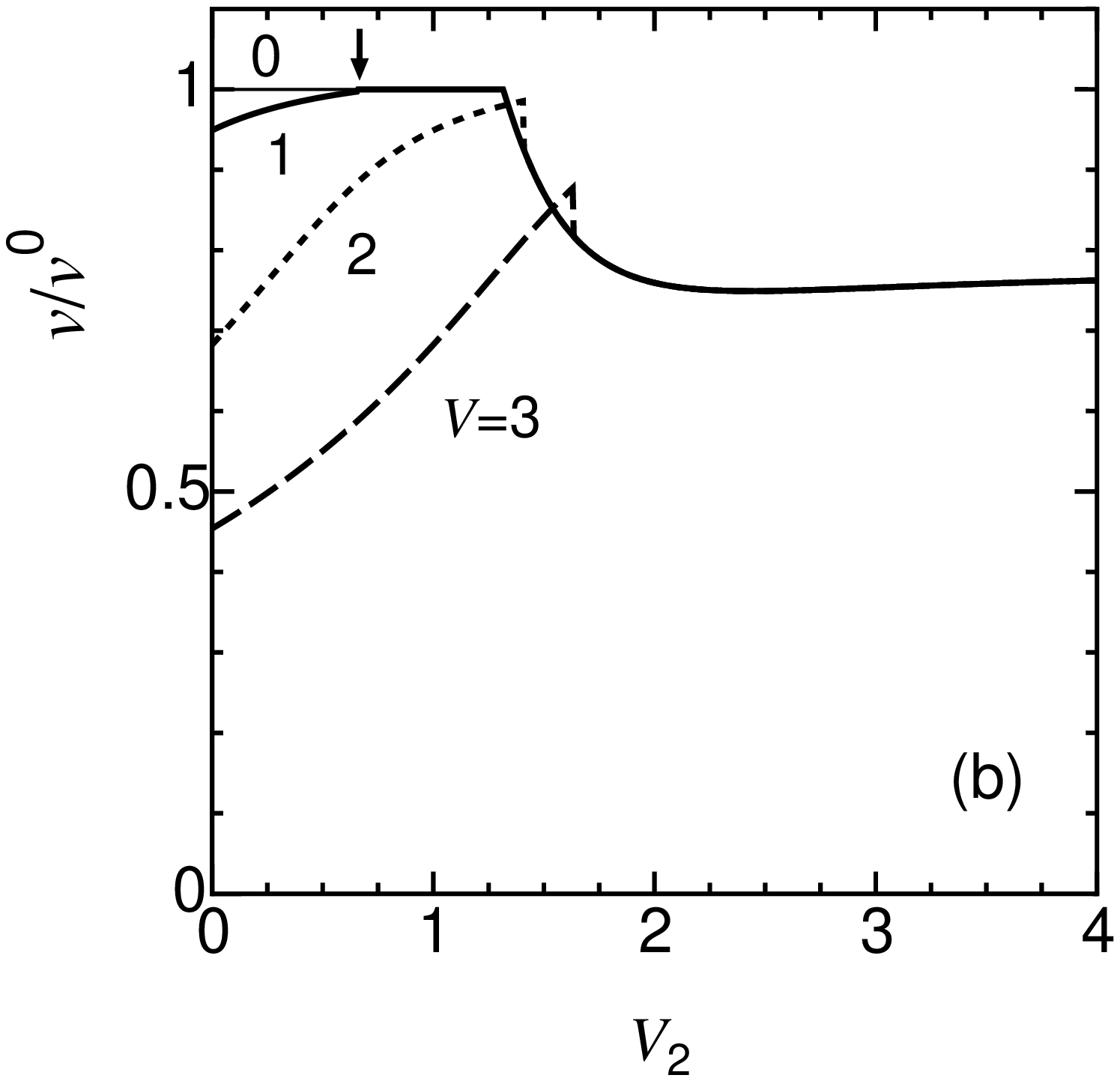}
\vspace*{2mm}
\caption[]{ (a)~
  $V_2$ dependence of $\chi$ for $V$=0 (Fig. 4), 1, 2 and 3.
There is a jump due to the first order transition between II and III
indicated by the dashed curve in Fig. 1. 
 The inset shows the $V_2$-dependence of $\rho$. 
~(b)~
 $V_2$-dependence of $v$ for $V$=0 (Fig. 4), 1, 2 and 3. 
For $V$=1, there is a small jump at $V_2$=0.66, which corresponds to
the transition between II and I. This discontinuity can also be seen
in Fig. 3. 
}
\end{center}
\end{minipage}
\end{figure}
%
\begin{eqnarray} 
 \label{vlimit}
 v/v^0 &=& \left( \frac{\rho}{\chi} \right)^{1/2}/v^0 
 = (t-t_{\rm d})/v^0 \;\,.
\end{eqnarray}
For $U=4$,  $v^0=1.286$, so that $v/v^0 =$ 0.777. For $V_2/t =4$ and
13, the numerical calculation gives $v/v^0 =$ 0.763 and  0.776,
respectively, in excellent agreement with the analytical result of the
two-site problem. 

Now we consider the $V_2$-dependence of $\chi/\chi^0$, $\rho/\rho^0$
and $v/v^0$ for $V =$ 0, 1, 2 and 3. For $V=1$, there is first a transition
from region II to region I, and then a transition from I to III. For
$V=2$ or 3, there is a single transition occuring between II and
III. The ratio $\chi/\chi^0$ and $\rho/\rho^0$ (inset) exhibit a similar
behavior (Fig.~5(a)). They are constant in region I, and increase
(decrease) in II (III) when $V_2$ increases. Figure 5(b) shows the
spin-wave velocity $v/v^0$ which turns out to be mainly determined by
$\rho/\rho^0$. Except in region I and for large values of $V_2$, $v$
varies strongly as a function of $V_2$. 

Here we  comment on the fact that $v$ remains finite at large $V_2$. 
 Within the mean-field treatment, which is expected to
 be valid for a moderate coupling between chains, both $\chi$ and
 $\rho$ remain finite at large $V_2$. On the other hand, for
 one-dimensional systems it is known from bosonization that $\chi$
 vanishes at large $V_2$ due to the formation of a spin
 gap. \cite{YoshiokaCREST} 
 Thus, we expect our mean-field analysis in
 region III of Fig.~1 to break down when the interchain coupling
 becomes sufficiently small.     
 
\vspace*{-1mm}
\begin{figure}[b]
\begin{minipage}{8cm} 
\begin{center}
\vspace*{0mm}
\leavevmode
\epsfysize=7.4cm
   \epsffile{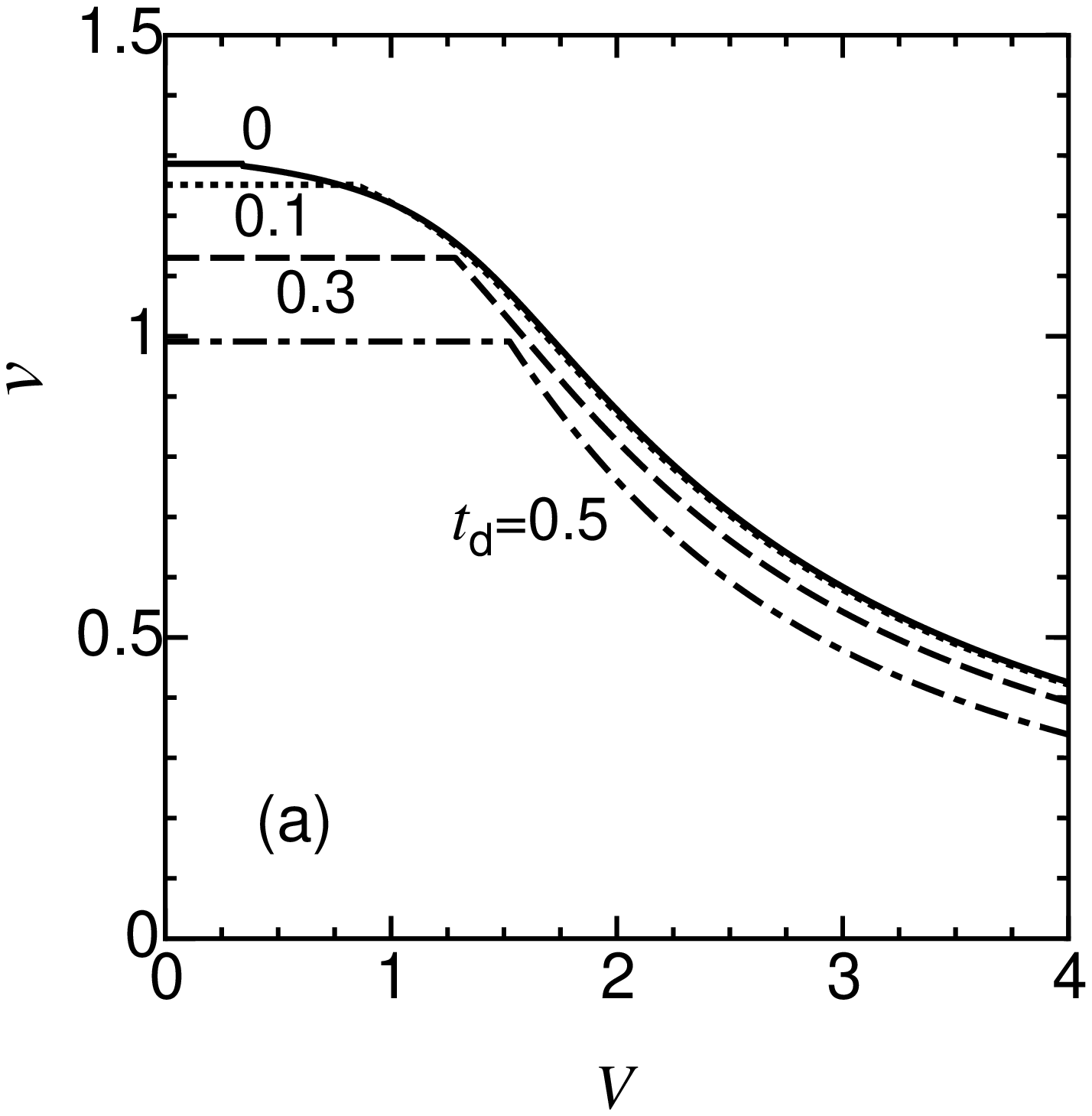}
\vspace*{3mm}
\leavevmode
\epsfysize=7.4cm
   \epsffile{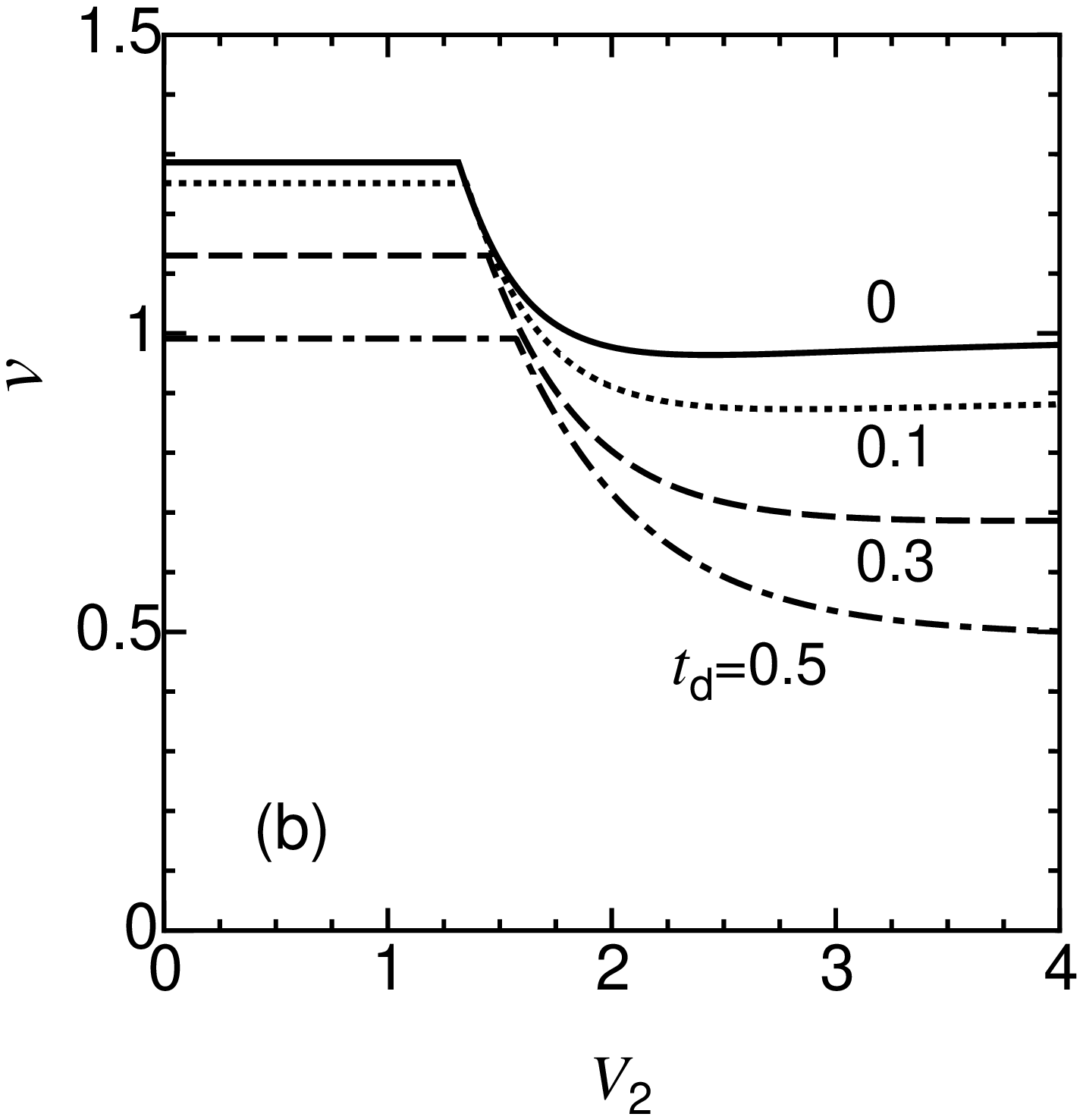}
\vspace*{2mm}
\caption[]{ (a)~
 $V$ dependence of $v$ for $U=4$, $V_2=0$ and 
 $t_{\rm d}$=0, 0.1, 0.3 and 0.5.  
~(b)~
 $V_2$ dependence of $v$ for $U=4$, $V=0$ and
 $t_{\rm d}$=0, 0.1, 0.3 and 0.5. 
}
\end{center}
\end{minipage}
\end{figure}
%

\subsection{Effect of dimerization} 

Finally, we consider the effect of dimerization on the spin-wave 
velocity $v$. Figure 6(a) shows  the $V$-dependence for $U=4$, $V_2=0$   
and $t_{\rm d}=0$ (solid curve), 0.1 (dotted curve), 0.3 (dashed curve) and
0.5 (dash-dotted curve). The effect of dimerization is large in region
I, but rather small in region II. A finite $t_{\rm d}$ increases the band
gap. This induces a suppression of $\Pi_{j_x^x j_x^x}$ and $\rho$, and
leads ultimately to a reduction of the spin-wave velocity. We note that
the reduction of $S_1$ and $D_2$ in region II by the dimerization has
little effect on $v$, since  the dependence of $S_1$ and $D_2$  on
dimerization  is very small for  $V \lesssim 4$. 

Figure 6(b) shows the $V_2$-dependence of $v$ for $U=4$, $V=0$ and
$t_{\rm d}=0$ (solid curve), 0.1 (dotted curve), 0.3 (dashed curve) and 0.5
(dash-dotted curve). The effect of dimerization is noticeable in both
regions I and III. The limiting behavior for large $V_2$ is given by
Eq.~(\ref{vlimit}).  In that limit, the SDW exists for $U > 2 (t-t_{\rm d})$ 
and the spin-wave velocity $v$ depends only on $t$  and $t_{\rm d}$.

\section{Conclusion}
\setcounter{equation}{0}

In conclusion, the nearest and next-nearest neighbor interactions 
strongly affects the
spin-wave velocity in the intermediate coupling regime $U\sim 4t$. Our
main results are as follows. (i) In the pure SDW state (region I), the
spin-wave velocity $v$ is independent of the nearest ($V$) and
next-nearest ($V_2$) interaction (Fig.~3). (ii) For coexisting
$2k_{\rm F}$ SDW and $4k_{\rm F}$ CDW (region II), $v$ decreases (increases) as
a function of $V$ ($V_2$) [Figs.~3 and 5(b)]. It is slightly
discontinuous at the transition between I and II and vanishes (as well
as the spin stiffness) at large $V$ (Fig.~3). (iii) For coexisting
$2k_{\rm F}$ SDW, $2k_{\rm F}$ CDW and $4k_{\rm F}$ SDW (region III), $v$ is 
 suppressed by $V_2$. It tends to a finite value at large
$V_2$ [Figs.~4 and 5(b)]. (iv) The dimerization decreases the spin-wave
velocity [Figs.~6(a) and (b)].  

As discussed in Sec.~III.B, our approach is limited to the weak to
intermediate coupling regime and should hold when $U,V,V_2 \lesssim
4t$. In the half-filled Hubbard model, a strong coupling is known to
reduce the spin-wave velocity from $v=O(t)$ to $v=O(J)$ (with
$J=4t^2/U\ll t$). We also expect a decrease of the spin-wave velocity
in the more general case we have studied when $U,V,V_2$ become larger
than $4t$. Therefore, our main conclusion (a reduction of the
spin-wave velocity by the interactions $V,V_2$) is likely to be strengthened
by strong coupling effects. The Stoner factor $(1-2\chi U)^{1/2}$, 
which arises from the coupling to long-wavelength spin fluctuations,
was not considered in our analysis. It leads to a decrease of
$v$ when the on-site interaction $U$ increases. Whether the Stoner
factor depends on the interactions $V$ and $V_2$ remains an open
question. 

In the compounds that have been studied experimentally,
\cite{Pouget,Kagoshima} the Bechgaard
salts (TMTSF)$_2$PF$_6$ and (TMTSF)$_2$AsF$_6$, and the
Fabre salt (TMTTF)$_2$Br, the electron-electron interaction is
expected to be in the intermediate coupling regime ($U\sim
4t$). Furthermore, estimates by Mila\cite{Mila} and quantum-chemistry
calculations\cite{Castet} have revealed the finite-range part of the Coulomb
potential, the first-neighbor interaction $V$ being equal or even larger than
$U/2$. We therefore think that our conclusions are relevant to the
Bechgaard-Fabre salts studied in Refs.~\onlinecite{Pouget} 
and \onlinecite{Kagoshima}.

\section*{ Acknowledgments }

 We (Y.T. and Y.S.) thank H. Sakanaka  for useful discussion.
This work was financially   supported by 
 Universit{\'e} Paris--Sud,   France and 
 a Grant-in-Aid 
 for Scientific  Research  from the Ministry of Education, 
Science, Sports and Culture (Grant No.09640429), Japan. 
 It was also supported by Core Research for Evolutional Science 
and Technology (CREST), Japan Science and Technology Corporation (JST).

\bleq
\appendix
\section{Derivation of eq.~(3.22)}
We rewrite Eq.~(\ref{Seff256}) as 
\begin{eqnarray}
 \label{Seff264}
  {\cal S}_{\rm eff} &=& \frac{1}{2} \int d\xi d\tau 
\left\{ 
  \chi (\partial_{\tau} {\bf n} )^2 
  + \rho (\partial_{\xi} {\bf n} )^2 
\right\}  
 - {}  \frac{1}{2} \sum_{\tilde{q}} 
\left\{
 \left( \left< K \right>_{\rm MF} + \Pi_{j_{x}^{z} j_{x}^{z}} \right) 
 |A_{x}^{z}(\tilde{q})|^2 
 + \Pi_{j_{0}^{z} j_{0}^{z}}|A_{0}^{z}(\tilde{q})|^2 
\right\}  \;\;,
\end{eqnarray}
where $\rho$ and $\chi$ are given by  
\begin{eqnarray}
 \rho = -\frac{1}{4} 
   \left( \left< K \right>_{\rm MF} + \Pi_{j_{x}^{x} j_{x}^{x}} \right)
          = -\frac{1}{4} 
   \left( \left< K \right>_{\rm MF} + \Pi_{j_{x}^{y} j_{x}^{y}} \right)
\;\;, 
\end{eqnarray}
and 
\begin{eqnarray}
 \chi = \left< S_{\nu} S_{\nu} \right>_{\rm MF} 
      = \frac{1}{4} \Pi_{j_{0}^{\nu} j_{0}^{\nu}} \;\;, 
\hspace{1cm} \nu=x,y \;\,.
\end{eqnarray}
To show that the second term of Eq.~(\ref{Seff264}) vanishes, we use
 the invariance of the action under the gauge transformation 
$
 A_{\mu}^{z}(\xi,\tau) \longrightarrow 
  A_{\mu}^{z}(\xi,\tau) 
 + \frac{1}{2} \partial_{\mu} \Lambda(\xi,\tau)  
$
 ($\mu=\xi$ or $\tau$) \cite{Schakel}. This transformation
 corresponds to a rotation of
 ${\bf n}_{\rm MF} = \hat{z}$ around the $\hat{z}$ axis and does not
 change the state of the system. The invariance of the action in this
 gauge transformation implies
\begin{eqnarray}
\label{gauge_trans}
 -\frac{1}{2} \sum_{\tilde{q}} 
\biggl\{ 
 \left( \left< K \right>_{\rm MF} + \Pi_{j_{x}^{z} j_{x}^{z}} \right)
 \left[ \frac{1}{4} q_x^2 |\Lambda(\tilde{q})|^2 
       - i A_{x}^{z}(\tilde{q}) q_x \Lambda(-\tilde{q})  \right]
 + \Pi_{j_{0}^{z} j_{0}^{z}} 
 \left[ \frac{1}{4}\Omega^2 |\Lambda(\tilde{q})|^2 
  + i A_{0}^{z}(\tilde{q}) \Omega \Lambda(-\tilde{q}) \right]
\biggr\} = 0 \;\,.
\end{eqnarray} 
\eleq
Since Eq.~(\ref{gauge_trans}) should be valid for an arbitrary
function $\Lambda$, we deduce  
\begin{eqnarray}
 \label{gauge1a}
 \left< K \right>_{\rm MF} + \Pi_{j_{x}^{z} j_{x}^{z}} &=& 0 \;\;, \\
 \label{gauge1b}
 \Pi_{j_0^z j_0^z} &=& 0 \;\;,
\end{eqnarray}
which lead to the vanishing of the second line of Eq.~(\ref{Seff264}). 

Equations (\ref{gauge1a}) and (\ref{gauge1b}) can also be obtained
from the U(1) electromagnetic field gauge invariance. 
Noting that $\Pi_{j_\mu^zj_\mu^z}=\Pi_{j_\mu^0j_\mu^0}$,
Eqs.~(\ref{gauge1a}) and (\ref{gauge1b}) can be rewritten as
\begin{eqnarray}
 \label{gauge2}
\left< K \right>_{\rm MF} + \Pi_{j_{x}^{0} j_{x}^{0}} &=& 0 \;\;, \\
 \label{gauge2b}
 \Pi_{j_0^0 j_0^0} &=& 0 \;\,.
\end{eqnarray}
We recognize here the components of the polarization tensor for the
usual U(1) electromagnetic gauge field. Equations (\ref{gauge2})
 and (\ref{gauge2b}) 
 follow from (electromagnetic) gauge invariance. \cite{Fradkin}


\section{Limiting case of large $V_2$}

When  $V_2 \rightarrow \infty$, the mean-field solution in region III 
corresponds to that of a half-filled two-site system given by 
\begin{eqnarray} \label{HB1}
 H &=& -(t-t_{\rm d}) \sum_{\sigma} 
           \left( C_{1\sigma}^{\dagger}C_{2\sigma} + {\rm H.c.} \right)
\nonumber \\
 & & {}
   + U \left( n_{1\uparrow} n_{1\downarrow} + 
                  n_{2\uparrow} n_{2\downarrow} \right)
    + V \sum_{\sigma,\sigma'} n_{1\sigma} n_{2\sigma'} \;\;,
\end{eqnarray}
where $n_{1\sigma}\,(n_{2\sigma})=
C_{1\sigma}^{\dagger}C_{1\sigma}\,(C_{2\sigma}^{\dagger}C_{2\sigma})$ 
 and $C_{1\sigma}^{\dagger}\,(C_{2\sigma}^{\dagger})$ denote 
 the creation operators of an electron at site 1\,(2) 
with spin $\sigma$.  The mean-field equations are given by
\begin{eqnarray} \label{OPB2}
   \sum_{\sigma} \left< C_{1\sigma}^{\dagger}C_{1\sigma}\right> 
 = \sum_{\sigma} \left< C_{2\sigma}^{\dagger}C_{2\sigma}\right> 
 &=& 1  \;\;, \\
                 \label{OPB3}
   \sum_{\sigma} \left< C_{1\sigma}^{\dagger}C_{1\sigma}\right> 
  {\rm sgn}(\sigma) &=&  \Delta \;\;, \\
                 \label{OPB4}
   \sum_{\sigma} \left< C_{2\sigma}^{\dagger}C_{2\sigma}\right> 
  {\rm sgn}(\sigma) &=&  -\Delta \;\;, 
\end{eqnarray}
 where the average $\left< \;\; \right>$ 
  is performed with the mean-field Hamiltonian
\begin{eqnarray} \label{HMFB5}
 H_{\rm MF} &=& -(t-t_{\rm d}) \sum_{\sigma} 
           \left( C_{1\sigma}^{\dagger}C_{2\sigma} + h.c. \right)
\nonumber \\
 & & {} + \sum_{\sigma} \left[ 
     \left( \frac{U}{2}+V-{\rm sgn}(\sigma)\frac{U}{2}\Delta \right)
       C_{1\sigma}^{\dagger}C_{1\sigma}
\right. \nonumber \\
 & & \left. {}
   + \left( \frac{U}{2}+V+{\rm sgn}(\sigma)\frac{U}{2}\Delta \right)
       C_{2\sigma}^{\dagger}C_{2\sigma} \right]
\nonumber \\
 & & {} -\frac{U}{2}-V+\frac{U}{2}\Delta^2  \;\,. 
\end{eqnarray}
From Eqs.~(\ref{OPB2}), (\ref{OPB3}), (\ref{OPB4}) and 
 (\ref{HMFB5}), 
the self-consistency equation for $\Delta$ is expressed as 
\begin{eqnarray} \label{SEB6}
 1 &=& \frac{U/2}{
 \sqrt{[(U/2)\Delta]^2+(t-t_{\rm d})^2} 
}
\;\;, 
\end{eqnarray}
where 
 $\mu=U/2+V$ at half-filling.  
The solution of Eq.~(\ref{SEB6}) is obtained as 
\begin{eqnarray} \label{solB7}
 \Delta &=& 
 \pm \sqrt{1-\left(\frac{2(t-t_{\rm d})}{U} \right)^2} \;\;,  
\end{eqnarray}  
 for $ U/(t-t_{\rm d}) > 2$.  
By using Eq.~(\ref{solB7}), we compute  
the uniform transverse spin susceptibility ($\chi'$) 
 and the spin stiffness ($\rho'$): 
\begin{eqnarray} 
\label{chiB8} 
 \chi' &\equiv& 
\frac{1}{2} \sum_{n,n'=1,2} 
 \left[ 
   \frac{1}{4}\left< j^x_0(n) j^x_0(n') \right> \big| _{i \Omega=0}  
 \right]
\nonumber \\
  &=& \frac{1}{2U}-  \frac{2(t-t_{\rm d})^2}{U^3} \;\;, \\
\label{rhoB9} 
 \rho' &=& - \frac{1}{4} 
        \left( \left< K' \right> + \Pi_{j^x_xj^x_x}' \right)
\nonumber \\
  &=& (t-t_{\rm d})^2 \left( \frac{1}{2U}-  \frac{2(t-t_{\rm d})^2}{U^3} \right)
 \;\;,
\end{eqnarray}    
where the kinetic energy per site ($\left< K'\right>$) and 
the spin current-current correlation function ($\Pi_{j^x_xj^x_x}'$)  
 are given by 
\begin{eqnarray}
\label{KB10} 
 \left< K' \right> &\equiv& 
\frac{1}{2} \left< -(t-t_{\rm d}) \sum_{\sigma} 
\left( C_{1\sigma}^{\dagger}C_{2\sigma} + {\rm H.c.} \right) \right>
\nonumber \\
 &=& - \frac{2(t-t_{\rm d})^2}{U} \;\;, \\
\label{PIB11} 
 \Pi_{j^x_xj^x_x}' &\equiv& 
\frac{1}{2} \sum_{n,n'=1,2}
   \left< j^x_x(n) j^x_x(n') \right> \big| _{i \Omega=0}
\nonumber \\
  &=& \frac{8(t-t_{\rm d})^4}{U^3} \;\;, 
\end{eqnarray} 
 with
\begin{eqnarray}
 \label{j0B12}
  j^x_0 (n) &=& \sum_{\sigma} C_{n,\sigma}^{\dagger} C_{n,-\sigma}   
 \;\;, \\
 \label{jxB13}
  j^x_x (n) &=& -\frac{i (t-t_{\rm d})}{2} \sum_{\sigma} \left(
     C_{1,\sigma}^{\dagger} C_{2,-\sigma}   
   - C_{2,\sigma}^{\dagger} C_{1,-\sigma} \right)  \;. 
\end{eqnarray} 
By noting that $\chi=\chi'/2$ and $\rho=\rho'/2$, we obtain the
spin-wave velocity of the one-dimensional system [Eq.~(\ref{H})] in
the limit $V_2\to\infty$ as
\begin{eqnarray} \label{vB14}
 v &=& \left( \frac{\rho}{\chi} \right)^{1/2} = t-t_{\rm d} 
\;\,.
\end{eqnarray}



\ecols


\begin{thebibliography}{99}
\bibitem{note} Fluctuations of $\Delta_s$ at wave vectors around $Q_0$
correspond to gapped amplitude fluctuations which do not couple to the
transverse spin-wave modes. (When two SDW's coexist in the
ground-state, there are both in-phase and out-of-phase amplitude
fluctuations.)
\bibitem{Berry} We neglect here the Berry phase term $S_{\rm
Berry}=-\sum_n \int d\tau \langle j_{0n}^z\rangle A^z_{0n}$ which is
expected to be irrelevant when interchain coupling is taken into
account.
\end{thebibliography}
\end{document}